\documentclass[%
 aip,
 amsmath,amssymb,
 reprint,%
]{revtex4-1}

\usepackage{graphicx}
\usepackage{dcolumn}
\usepackage{bm}

\usepackage[utf8]{inputenc}
\usepackage[T1]{fontenc}
\usepackage{mathptmx}
\usepackage{etoolbox}
\usepackage{xcolor}
\makeatletter
\def\@email#1#2{%
 \endgroup
 \patchcmd{\titleblock@produce}
  {\frontmatter@RRAPformat}
  {\frontmatter@RRAPformat{\produce@RRAP{*#1\href{mailto:#2}{#2}}}\frontmatter@RRAPformat}
  {}{}
}%
\makeatother

\usepackage{lineno}

\begin{document}

\preprint{APS/123-QED}

\title{An apparatus for in-vacuum loading of nanoparticles into an optical trap}

\author{Evan Weisman}
\author{Chethn Krishna Galla}
\author{Cris Montoya}
\author{Eduardo Alejandro}
\affiliation{
 Center for Fundamental Physics, Department of Physics and Astronomy, Northwestern University, Evanston, IL 60208
}%
\author{Jason Lim}
\author{Melanie Beck}
\affiliation{Department of Physics, University of Nevada, Reno, NV 89557}
\author{George P. Winstone}
\author{Alexey Grinin}
\author{William Eom}
\author{Andrew A. Geraci}
\affiliation{
 Center for Fundamental Physics, Department of Physics and Astronomy, Northwestern University, Evanston, IL 60208
}%


\date{\today}

\begin{abstract}
We describe the design, construction, and operation of an apparatus utilizing a piezoelectric transducer for in-vacuum loading of nanoparticles into an optical trap for use in levitated optomechanics experiments. In contrast to commonly used nebulizer-based trap-loading methods which generate aerosolized liquid droplets containing nanoparticles, the method produces dry aerosols of both spherical and high-aspect ratio particles ranging in size by approximately two orders of mangitude. The device has been shown to generate accelerations of order $10^7$ $g$, which is sufficient to overcome stiction forces between glass nanoparticles and a glass substrate for particles as small as $170$ nm diameter. Particles with sizes ranging from $170$ nm to $\sim 10$ $\mu$m have been successfully loaded into optical traps at pressures ranging from $1$ bar to $0.6$ mbar. We report the velocity distribution of the particles launched from the substrate 
and our results indicate promise for direct loading into ultra-high-vacuum with sufficient laser feedback cooling. This loading technique could be useful for the development of compact fieldable sensors based on optically levitated nanoparticles as well as matter-wave interference experiments with ultra-cold nano-objects which rely on multiple repeated free-fall measurements and thus require rapid trap re-loading in high vacuum conditions.
\end{abstract}

\maketitle

\section{\label{sec:intro} Introduction}


Levitated nanoparticles have proven to be exceptional force sensors operating at high vacuum, exhibiting high quality factors \cite{ricci2017optically} and zeptonewton sensitivity \cite{zepto}. A range of nanoparticle diameters of 50 nm – 20 $\mu$m have been used in these systems \cite{Moore2021} making them ideal probes for studying surface forces \cite{Winstone2018,montoya2022scanning,Diehl2018}, searching for new short-range forces \cite{Geraci2010}, detecting gravitational waves \cite{Aggarwal2020}, and even detecting dark matter \cite{Monteiro2020}. Feedback cooling of nanoparticles to their ground state \cite{Delic2020} has been achieved showing much promise for the investigation of quantum phenomena at the macroscale. Experiments and applications such as matter-wave interference, which requires a coherent quantum state of a nanoparticle during free-fall, rely on superb environmental isolation to achieve quantum level sensitivity \cite{Bateman2014,Geraci2015,Rademacher_2019}. All these systems must operate at high vacuum, and a method for launching and trapping nanoparticles directly at ultra high vacuum is essential for fast re-loading particles into compact fieldable sensors based on optically levitated nanoparticles as well as matter-wave interference experiments with ultra-cold nano-objects which rely on multiple repeated free-fall measurements.

The vast majority of levitated optomechanics experiments have employed nebulizers for loading of particles into the optical trap. This method allows a wide variety of particle sizes to be loaded, in particular below $200$ nm diameter. The disadvantage of this method is that the particles need to be loaded at atmospheric pressure, requiring long pump-downs. Furthermore the introduction of solvents or moisture into the vacuum apparatus further complicates the process of achieving ultra-high vacuum.  Such methods are particularly poorly suited to experiments requiring repeated trapping of nanoparticles in vacuum, such as nanoparticle interferometry experiments, due to the excessively long duty cycle between measurements. Delivering particles from a dry substrate in vacuum has been investigated in several experiments, which has been a robust method for particle sizes exceeding 1 micron diameter \cite{Li:2011,Moore:2014, Moore:2021}. Some success has been noted, although is less reproducible, for shaking a drumhead \cite{Barkerprl2015}.  Other techniques that are compatible with vacuum include Laser-Induced Acoustic Desorption (LIAD), which has been used to load particles into ion traps \cite{bykov2019direct} as well as optical traps \cite{photonics8110458}.

In this paper, we describe an apparatus which uses a glass substrate clamped to a piezoelectric transducer for robust in-vacuum loading of dielectric nanoparticles into an optical trap. The device has been shown to generate accelerations of order $10^7$ $g$, which is sufficient to overcome stiction forces between glass nanoparticles and a glass substrate for particles as small as $170$ nm diameter, consistent with theoretical expectations. Using this device, single spherical particles with sizes ranging from $170$ nm to $3$ $\mu$m \cite{Atherton:2015,Ranjit:2016,montoya2022scanning}, as well as clumps of such particles \cite{Atherthesis}, have been successfully loaded into optical traps at pressures ranging from 1 bar to 0.6 mbar. We have also successfully trapped Yb-doped $\beta-$NaYF sub-wavelength-thickness high-aspect-ratio hexagonal prisms with a micron-scale radius \cite{winstone2022}. We report the velocity distribution for a variety of particles launched from the substrate, 
and our results indicate promise for direct loading into ultra-high-vacuum with sufficient laser feedback cooling. In contrast to other clean vacuum compatible loading methods such as LIAD, where particles tend to be launched with speeds well in excess of 1 m/s \cite{photonics8110458}, our method produces sufficient flux of particles with initial launch velocity less than 1 m/s to in principle allow direct ultra-high-vacuum trapping with the assistance of a slowing laser beam. 

\section{Stiction Forces}
Stiction is a phenomenon that has been known to limit the operation of microelectromechanical systems (MEMS), whereby nearby structures become affixed due to surface interactions. In our case, there are three
predominant stiction forces to be considered, each of which depends on the
separation distance between the objects \cite{stiction}, the Van der Waals force, the electrostatic double-layer force, and the capillary force. The latter two forces can be present if there is residual liquid at the interface between the surfaces of the materials.

The Van der Waals force is the sum of attractive or repulsive forces between molecules and consists of three distinct interactions. The first is the
Keesom force which describes the force between two permanent dipoles. The
Debye force details the force between a permanent dipole and an instantaneously induced dipole. Finally, the London dispersion force describes the force between two instantaneously induced dipoles. The latter is the most
important of the three because it is the only one that is always present regardless of the physical situation. In general, the total Van der Waals force between a sphere and a surface can be approximated as \cite{ISRAELACHVILI2011191, ISRAELACHVILI2011205, ISRAELACHVILI2011223} $F_{\mathrm{VDW}}=AR/6D^2$, where $A$ is the Hamaker constant, $R$ is the radius of the sphere, and $D$ is the separation distance between a sphere and a flat surface. Evaluating this expression for D = 0.4 nm and a sphere radius of $150$ nm, we arrive at an adhesion force between 6.25 nN and 62.5 nN for the Hamaker constant A = .4 - 4 10$^{-19} $J. Though
not considered explicitly in the previous equation, the Van der Waals force
is highly dependent on surface roughness. It has been demonstrated to drop
significantly for surface roughness greater than $1$-$2$ nm \cite{roughness}.

The electrostatic double layer force is associated with an object placed in
liquid. If the object possesses any inherent charge a double layer of ions
forms on its surface. The first layer is composed of ions adsorbed directly
onto the object by any of a number of chemical reactions resulting in a
wall of surface potential. The second layer consists of oppositely charged
free ions that electrically screen the first layer \cite{doublelayer}. Though our particles
are assumed neutral, if there is any net charge, this force could come to
dominate. However, due to the nature of dielectric material it is difficult to
measure.

The third adhesive force to consider is the capillary force. This force is
composed of the capillary pressure force and the surface tension force. In
the regime of interest (less than $1$ $\mu$m) the surface tension force is predominant. Furthermore the capillary force becomes highly dependent on ambient
humidity to the extent that the capillary force could vanish at low enough
humidity \cite{capillary}. 

In order to minimize the effects from double-layer or capillary forces, we dessicate the nanoparticles on a hotplate at 150$^{\circ}$C for a minimum of 15 minutes to remove excess moisture prior to loading them into the vacuum chamber.

Because of the difficulty in correctly measuring the individual values of
each of these interactions in order to properly determine the total adhesion
forces present one can instead estimate the ``pull-off'' force. Derjaguin, Muller and Toporov (DMT) \cite{DMT} proposed a model
which incorporates attractive forces outside but in the vicinity of the contact area. This model lends itself to small, hard bodies with low surface energy. The force is
\begin{equation}
    F_{\mathrm{DMT}}=4 \pi R_{\mathrm{eff}} \gamma
\end{equation}
where $\gamma$ is the effective surface energy and $R_{\mathrm{eff}}$ = $R R_{\mathrm{surf}} / (
R+R_{\mathrm{surf}})$ is the reduced radius of curvature for the microsphere and surface. It has
been experimentally determined that this model accurately describes the
proportionality between the pull-off force and microsphere radius for silica
particles with radii between $0.5$ to $2.5$ $\mu$m \cite{stiction}. Although the smallest particles we study are not within this previously tested range, we theoretically expect the model to be applicable and assume it to be valid in order to roughly estimate the necessary pull-off force for our system.  The acceleration to pull-off silica spheres from a glass substrate is about $ 10^6-10^8$ m/s$^2$ for 3 $\mu$m - $300$ nm diameter spheres \cite{stiction}. 

\section{\label{sec:device} Experimental Apparatus}

\begin{figure}[htbp]{}
 \includegraphics[width=1.0\linewidth]{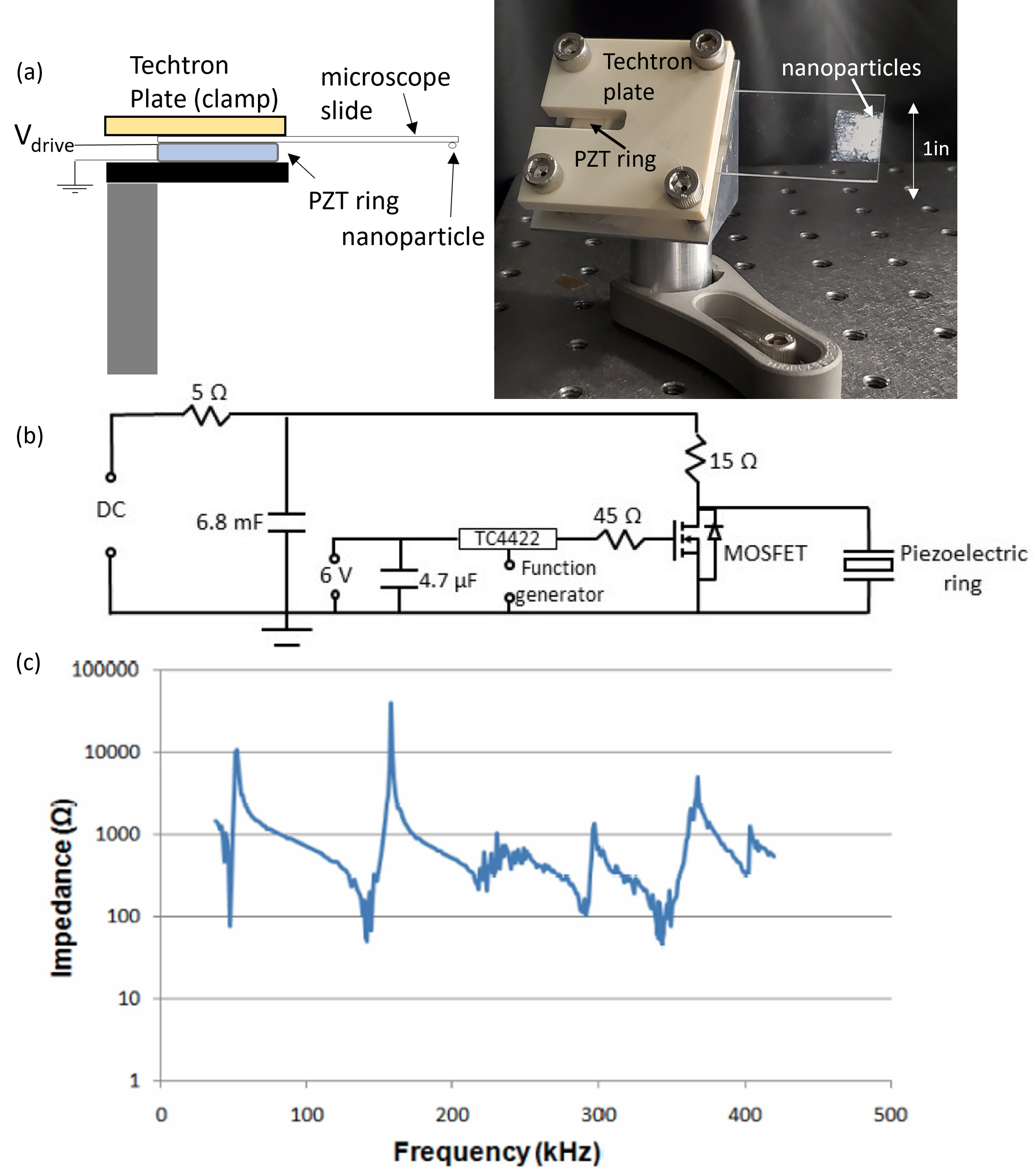}
 	\caption{(a) Left: Nanoparticle launcher consists of a clamped piezoelectric ring and microscope slide with deposited nanoparticles. When the piezoelectric ring is driven on resonance with high voltage, particles are shaken off of the slide, allowing them to fall into the optical trapping region. Right: 45${^\circ}$ angled launcher with nanoparticles coating the end of the slide. The angled launcher was chosen to reduce stray light scatter from the nanoparticle imaging setup. (b) Launcher circuit diagram. DC voltage up to 180 v is applied to the large capacitor which is grounded through the power MOSFET at the PZT resonance. (c) Impedance vs frequency of the PZT ring. Resonances occur at the impedance minima, notably near 140 and 340 kHz.     }
	\label{fig:device_circ_imp}
\end{figure}

The device that is developed for loading nanoparticles into the trap consists of a hard ceramic ring-shaped piezoelectric transducer (PZT), a clamping mechanism, a glass substrate, and a pulsed-power generator capable of generating sufficient current for driving the PZT at its resonance frequencies with large amplitudes, which shakes the glass substrate thus releasing the nanoparticles for trapping. 
The nanoparticles are prepared from a solution, which is pipetted onto glass substrates, and baked to extract as much moisture as possible. The dried particles are then smeared on the end of a glass microscope slide that is clamped with the piezoelectric ceramic piece between aluminum and Techtron plates, as depicted in Fig. \ref{fig:device_circ_imp} (a). Two sides of the piezoelectric ceramic are coated in silver to act as electrodes, and connected with Silver/Tin solder to the pulsed power generator.


The high-powered signal generator essentially consists of two simple sub-circuits: a power MOSFET connected to a high-voltage power supply with a large capacitor to store charge, and a smaller gate driving MOSFET connected to a function generator which provides the alternating signal. Fig. \ref{fig:device_circ_imp} (b) shows the general schematic for the driver.  A sense resistor is also connected in parallel (not shown) with the piezo output in order
to sample the signal delivered to the piezo. Since a very high current will be running through the sub-circuit containing
the power MOSFET, we also implement a very large copper heat sink onto which the power resistors and the MOSFET are mounted. 
The sense resistor is also mounted on its own smaller heat sink. Due to the fact that in a previous implementation of this design these electrical components kept overheating for a sufficiently large duty cycle, a DC fan has been implemented to further cool the components.

To achieve a maximal surface acceleration, the piezo is driven at a resonance frequency at which the input impedance will be very low. Due to this low input impedance, the driver must provide a very high current to the piezo electrodes in order to drive the piezo with a large amplitude. The impedance measured for the PZT is shown in Fig. \ref{fig:device_circ_imp} (c).  



\begin{figure}
    \includegraphics[width=\linewidth]{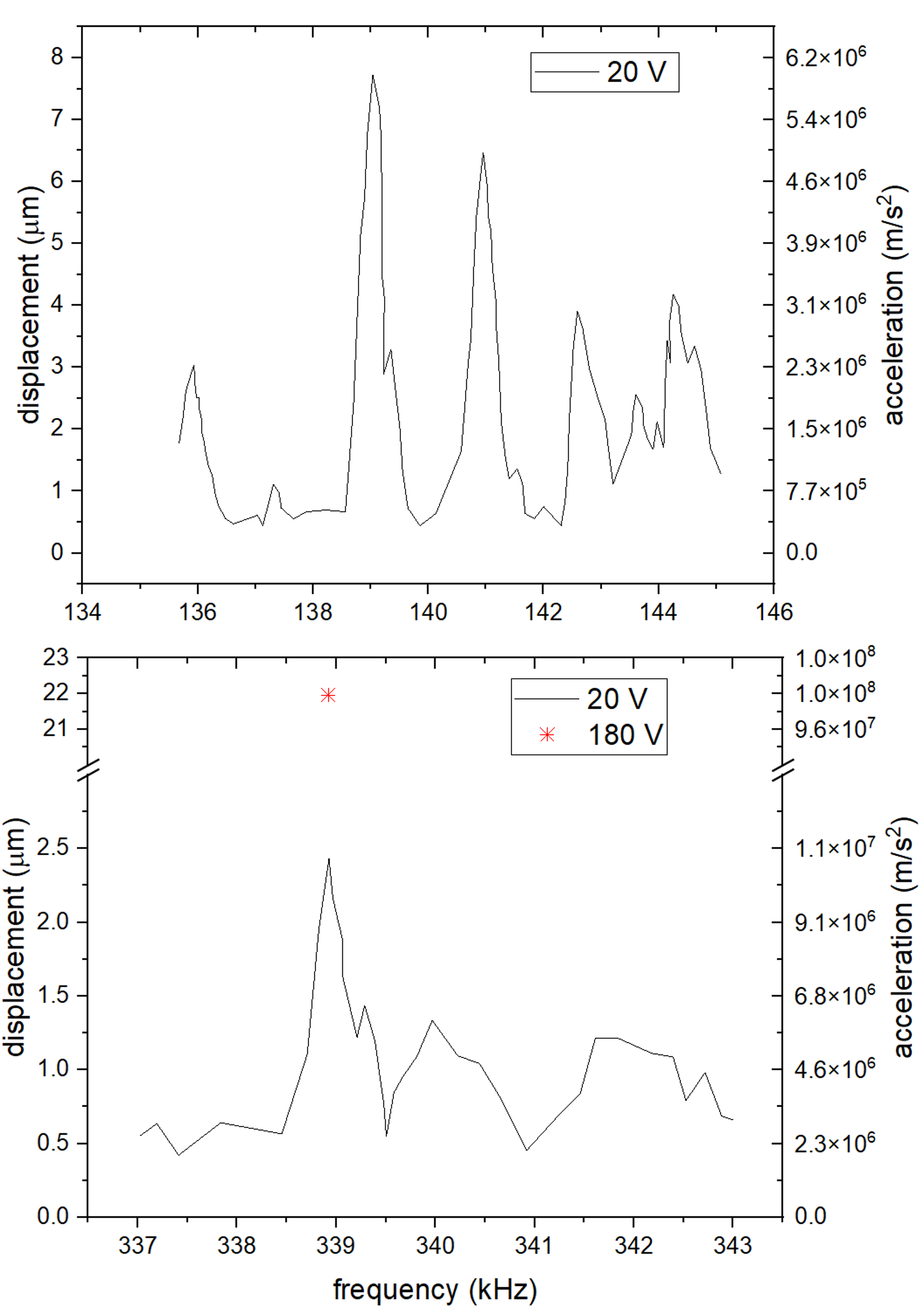}
	\caption{Displacement and acceleration spectra at driving frequencies of 140 kHz (top) and 340 kHz(bottom). The pulsed power generator was set to 20 V for these measurement, but can be set up to 180 V where accelerations of 10$^8$ m/s$^2$ are observed. The PZT was pulsed for 150 ms.}
	\label{fig:displacement}
\end{figure}

To measure the acceleration experienced by the nanoparticles
a helium-neon laser is reflected off of the slide which holds the nanoparticles.
The reflected beam is then directed to a quadrant photodetector
where the deflection of the beam due to the displacement of slide can be measured. The acceleration experienced by the particles is
approximately $a = \omega^2 \delta$, where $\omega$ is the angular frequency of displacement and $\delta$ is
the peak displacement near the end of the slide where the particles are deposited.  Fig. \ref{fig:displacement} shows the displacement and acceleration vs PZT frequency.  

\section{\label{sec:velocity} Velocity Measurements}
 We characterize our nanoparticle loading mechanism by measuring the velocity of falling nanoparticles of 3 $\mu$m diameter as a function of pressure as they are launched into an optical trap. The trap consists of dual counterpropagating 1064 nm beams focused down to 8 $\mu$m waists separated by about 75 $\mu$m as in \cite{Atherthesis,zepto}.
 In the experimental setup the particles are launched and trapped under moderate vacuum at 5-10 mbar using the launching method described in section \ref{sec:device}. The nanoparticle coated glass slide is clamped to the piezoelectric transducer on the mount tilted at $45^{\circ}$ as in  Fig. \ref{fig:device_circ_imp}(a) and vibrated near one of the resonances at approximately $140$ kHz or $339$ kHz, launching the nanoparticles towards the optical trap with its beam axis approximately 33 mm below the slide. To maximise particulate flux and efficiently deplete the slide, we find it necessary to systematically vary the driving frequency of the launcher in $\sim$ $\pm 500$ Hz steps around the central resonance at each driving voltage - before proceeding to increase the voltage. 
 
\begin{figure}
    \includegraphics[width=\linewidth]{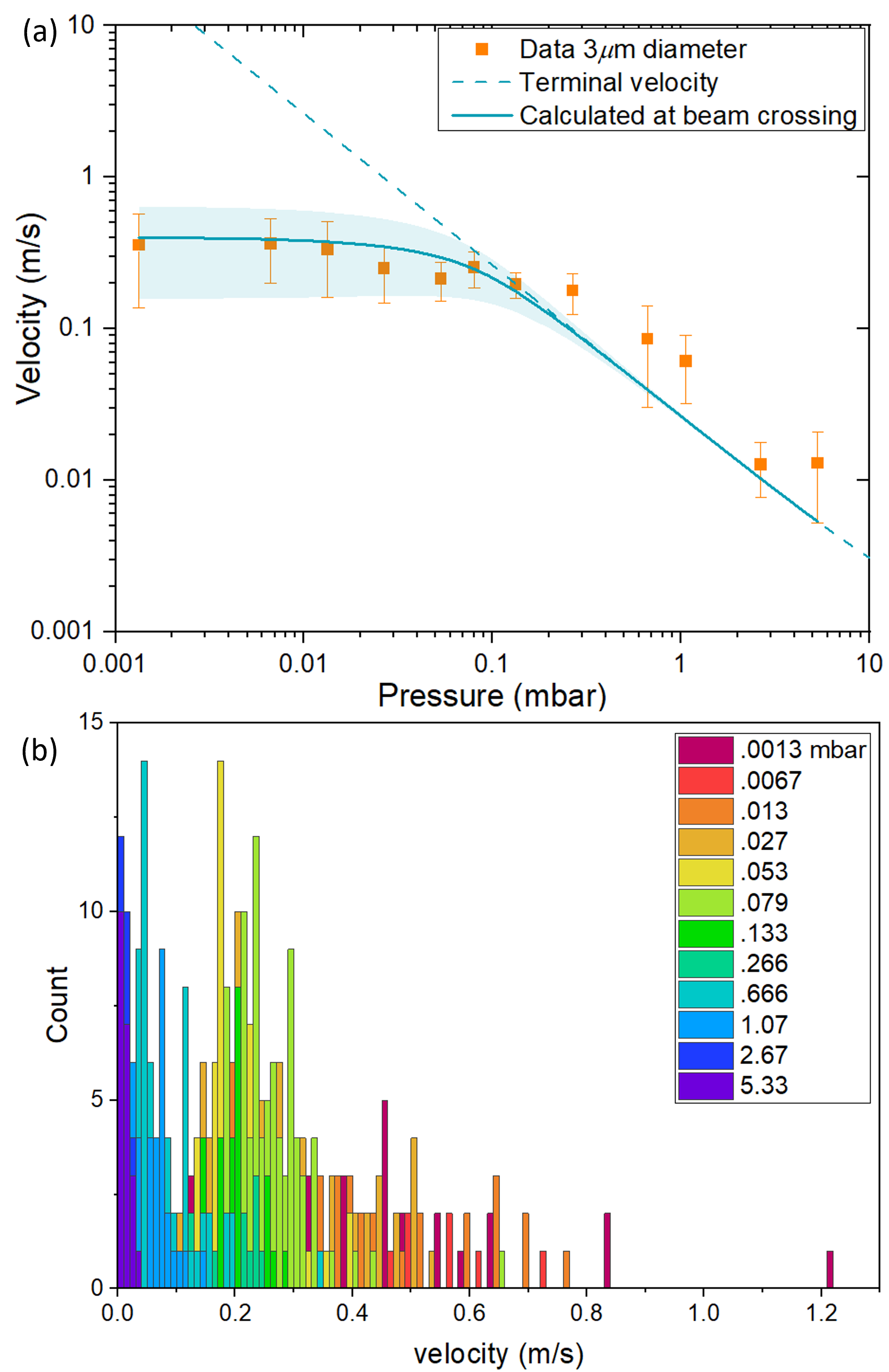}
	\caption{Average velocity of $3\mu$m diameter particles vs pressure released from a $45^{\circ}$ angled launcher. Solid line: calculated average velocity after falling $.033$ m. The spread in velocities at each pressure are primarily due to particle clumps at low vacuum and vibration time of the launcher at high vacuum. Band around the solid line is the estimated uncertainty in velocity due to the uncertainty in the time the particle is released from the substrate by the launcher mechanism.  (b) Histogram of the $3\mu$m particle average velocities. Colors indicate the particular pressure.}
	\label{fig:VvsP}
\end{figure}

 We synchronized a highspeed CCD camera to begin recording video at the start of the launch and capture the flux of particles through the counter-propagating beams. The time to beam crossing for each illuminated particle is $t_i = N_i/F_r$ , where $F_r$ is the camera's frame rate (up to 217/s) and $N_i$ is the number of frames from the beginning of launch to when particle $i$ scatters the trap light. The average falling velocity for each particle is then $v_i = $(33 mm)$F_r/N_i$. We measured falling rates at vacuum pressures of 5 to $10^{-3}$ mbar. 
 Our results are shown in Fig. \ref{fig:VvsP} (a) along with the projected terminal velocity and calculated average velocity at the beam crossing after falling 33 mm. Additional data taken for 300 nm diameter particles launched in the same system and 170 nm diameter particles launched in a system with a single beam trap can be found in the appendix. The velocity distribution histogram  of 3 $\mu$m diameter particles is shown in Fig. \ref{fig:VvsP} (b).
 
The spread in the velocity data at each pressure in Fig. \ref{fig:VvsP} (a), particularly for speeds greater than the terminal velocity at pressures above $10^{-1}$mbar, is likely due to the presence of clumped particles. Clumps of two or more particles are shaken off more easily due to less stiction force to the substrate \cite{stiction} and will achieve a higher terminal velocity, since $v_t \sim r$ for the pressure regime we study. We also expect clumps to appear in earlier frames as they scatter more light at the fringes of the optical trap. At higher vacuum, the vibration time of the launcher also contributes significantly to the uncertainty in velocity. The launcher was vibrated up to 200 ms during launch and assuming the particles leave on average after 100ms,  we estimate that the 1-sigma uncertainty in fall time due to only the launcher is $\pm$50 ms. Propagating the uncertainty in the fall time to the average particle velocity gives $\sigma_v = $ (.033 m)$\times(1/t^2)\times$(50 ms) for the particle fall time $t$. At high vacuum with the average velocity  $v=$ .400 m/s, $\sigma_v = $.245 m/s and at a moderate vacuum of .3 mbar with $v = .100$ m/s,  $\sigma_v = $.015 m/s. The uncertainty in the launcher timing contributes most significantly then to the spread in Fig. \ref{fig:VvsP} (a) for the fastest particles, which occur prominently at high vacuum.   
 
 The systematic uncertainty due to vibration time can be reduced by implementing a frame to frame account of velocity at high vacuum or a system with two or more stacked optical traps that work as gates. Alternatively one could employ a vertically oriented beam to illuminate the particle during its entire falling path from the substrate to the trap location (see Sec. \ref{sec:capture and slowing}). In principle, a camera with a frame rate of $217$ frames/s can capture 3 frames of an object as fast as $2.4$ m/s over the $.033$ m fall distance. In our setup we observe particles falling near the trap location as they are illuminated by the trapping lasers; for instance to measure velocities within $5$ mm of the trap our camera would capture 3 frames for particles only as fast as $.36$ m/s. So an accurate frame to frame measure of velocities would require an even higher speed camera than the one used in this setup.


Nanoparticles falling in moderate to high vacuum experience a drag force linear in velocity, their equation of motion is given by  \begin{equation}\label{fall} mg- m\gamma \dot{x} = m\ddot{x} .\end{equation} Here $m$ is the mass of the particle, $g$ is the acceleration due to gravity and $\gamma$ is the drag coefficient or damping rate as the nanoparticle collides with air molecules. The damping rate described by kinetic theory is  \begin{equation}\label{gamma} \gamma = \frac{6\pi \eta r }{m}\frac{.619}{.619+K_n}(1+c_k) .\end{equation} The flow of nanoparticles through rarefied gas is characterized by Knudsen's number $K_n = l/r$, where $l$ is the mean free path length and $r$ is the radius of the nanoparticle. At low pressure we have $K_n >> 1 $, the regime of flow where the mean free path is much larger than the particle size. The viscosity of the gas is $\eta$, and $c_k = (.31 K_n)/(.785 + 1.152 K_n + K_n^2)$. Since the mean free path is inversely proportional to pressure, $\gamma$ scales linearly with pressure in low pressure \cite{beresnev_chernyak_fomyagin_1990}. 
Setting the total force on the particle to zero in Eq. \ref{fall} gives the terminal velocity $v_t = g/\gamma$. We found that below $10^{-1}$ mbar, the particles don't reach the terminal velocity and fall 33 mm in about 82 ms. To calculate the velocity as a function of pressure we find the time it takes the particles to reach 33 mm by solving Eq. \ref{fall} for position,
\begin{equation} \label{x} x(t) =v_t(t+\frac{v_t}{g})\left[e^{-tg/v_t}-1\right]-v_i \frac{v_t}{g}(e^{-tg/v_t}-1),\end{equation}  
 and finding the time $t_f$ for $x(t_f) = $ 33 mm.  
 $x(t_f)/t_f$ is the average velocity 'calculated at beam crossing' curve plotted in Fig. \ref{fig:VvsP} vs pressure. 
 
 Despite the large accelerations of our ultrasonic launcher, nanoparticles were observed to fall with zero initial velocity and we set $v_i = 0$ in the above calculations. At high vacuum it is possible that the $45^{\circ}$ angled launcher ejects fast particles away from the optical trap and are unobserved in these measurements. Spheres launched with zero initial velocity are indicative of the launch acceleration just overcoming the sphere-to-substrate stiction forces.
 
\section{\label{sec:capture and slowing} Slowing beam for high vacuum trapping}

\begin{figure}
    \includegraphics[width=\linewidth]{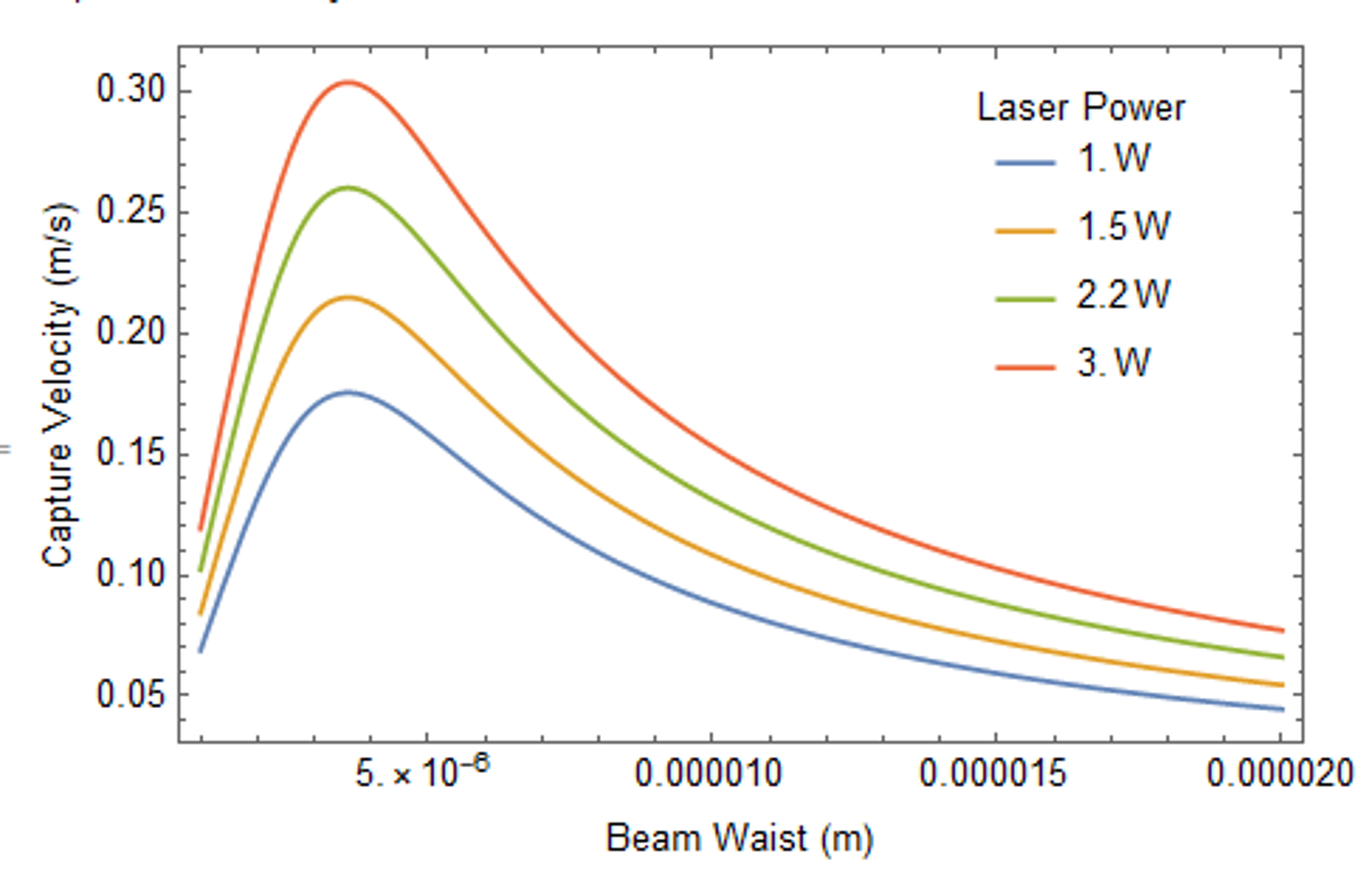}
	\caption{Maximum capture velocities for a range of beam waists and laser powers at the center of a counter propagating optical trap. The gap between foci is $75 \mu$m.}
	\label{fig:capvs}
\end{figure}

A nanoparticle will be captured by the optical trap, consisting of the conservative potential in equation \ref{trappot_r_z}, as long as its kinetic energy at the potential minimum is reduced below the trap’s potential depth by sufficient gas damping or laser scattering. At high vacuum laser scattering would be necessary to slow the particle for trapping. Immediately upon trapping, laser feedback cooling is then required to stabilize the particle against the non-conservative radiometric and scattering forces, and forces due to laser technical noise to keep the particle trapped at high vacuum \cite{Atherton:2015} . The potential for a small ($r << \lambda$) spherical particle of volume V in an orthogonally polarized counter propagating dual beam trap is given by
\begin{equation}\label{trappot_r_z}
U(r,z) = -3V\frac{(\epsilon-1)}{(\epsilon+2)}\frac{Iw_0}{2c}\left[ \frac{e^{-2\frac{r^2}{w(z)^2}}}{w(z)^2}
+\frac{e^{-2\frac{r^2}{w(z-z_g)^2}}}{w(z-z_g)^2}\right],
\end{equation} with the waist $w(z) = w_0\sqrt{1+(z/z_r)^2}$.
At the trap center

\begin{equation}\label{trapPotential} U(r=0,z=\frac{z_g}{2}) = -3 V \frac{(\epsilon-1)}{(\epsilon+2)}\frac{I}{c}\left[1+\left(\frac{z_g/2}{z_r}\right)^2\right]^{-1}.
\end{equation}

$I$ is the total trap intensity, $z_g$ is gap between foci and $z_r =\pi \omega_0^2/\lambda$ is the Rayleigh range.
\begin{equation}
|U| = \frac{1}{2} m v^2 
\end{equation} gives the escape velocity 
\begin{equation}
 v = \sqrt{\frac{6}{\rho} \frac{(\epsilon-1)}{(\epsilon+2)}\frac{I}{c}\left[1+\left(\frac{z_g/2}{z_r}\right)^2\right]^{-1} }
\end{equation} for a particle of density $\rho$ and dielectric constant $\epsilon$ \cite{Atherthesis}.

 \begin{figure}
    \includegraphics[width=\linewidth]{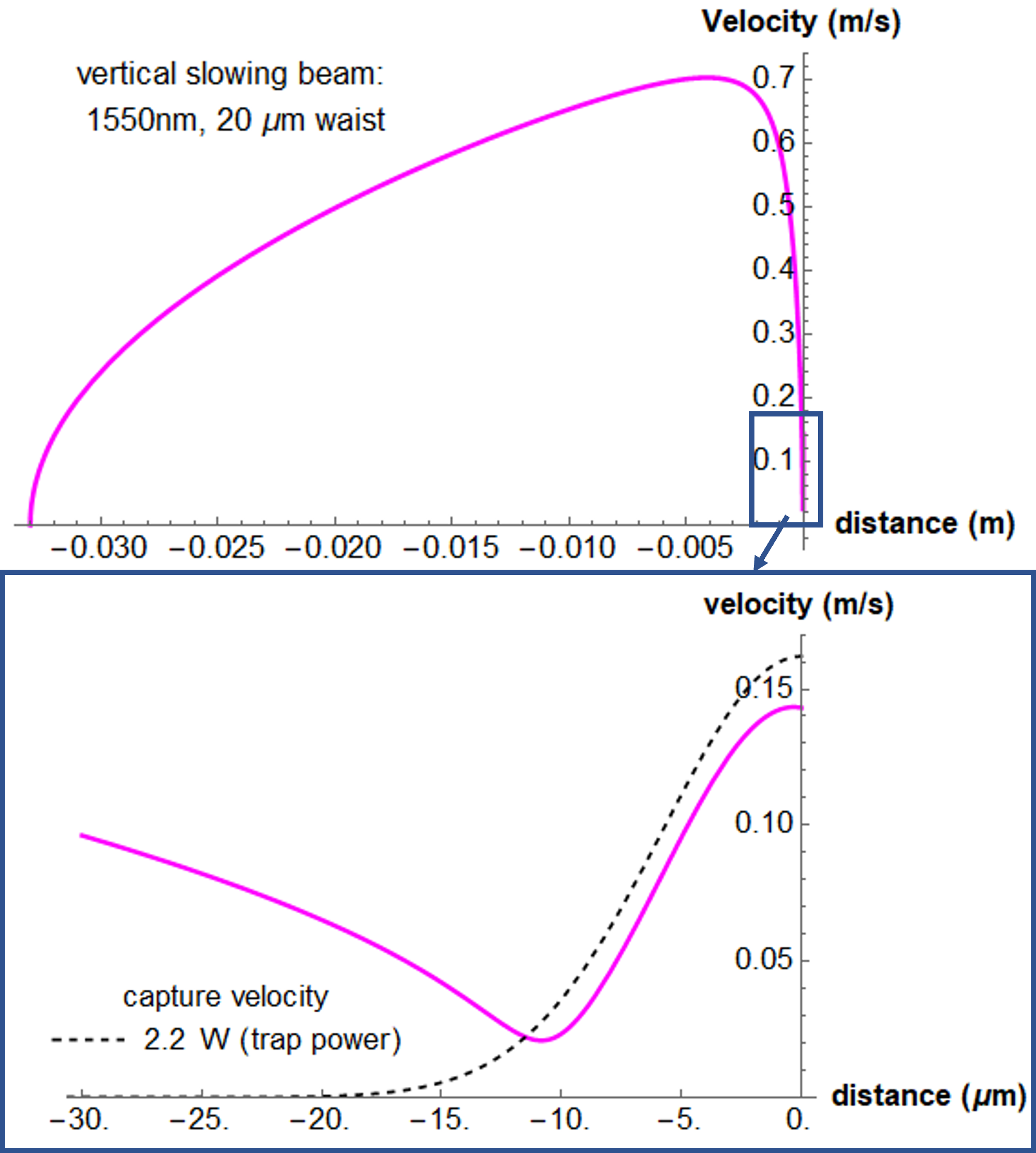}
	\caption{Velocities of 170 nm, 300 nm and 3 $\mu$m particles falling from 0.033 m into a counterpropagating dual beam optical trap with 8 $\mu$m waists. Top: Particle's velocity is initially gained due to free-fall and then slowed dramatically as it approaches the focus of the vertical slowing beam at the center of the trap. Bottom: Zoomed in around 30 $\mu$m of the trap. Trapping occurs for velocities below the capture velocity shown by the black dashed line. The slowing beam power for each sphere diameter is adjusted to achieve the same slowing trajectory: 170 nm at 4.32 W, 300 nm at .79 W, and 3 $\mu$m at .448 W}
	\label{fig:slow}
\end{figure}

\begin{figure}
    \includegraphics[width=\linewidth]{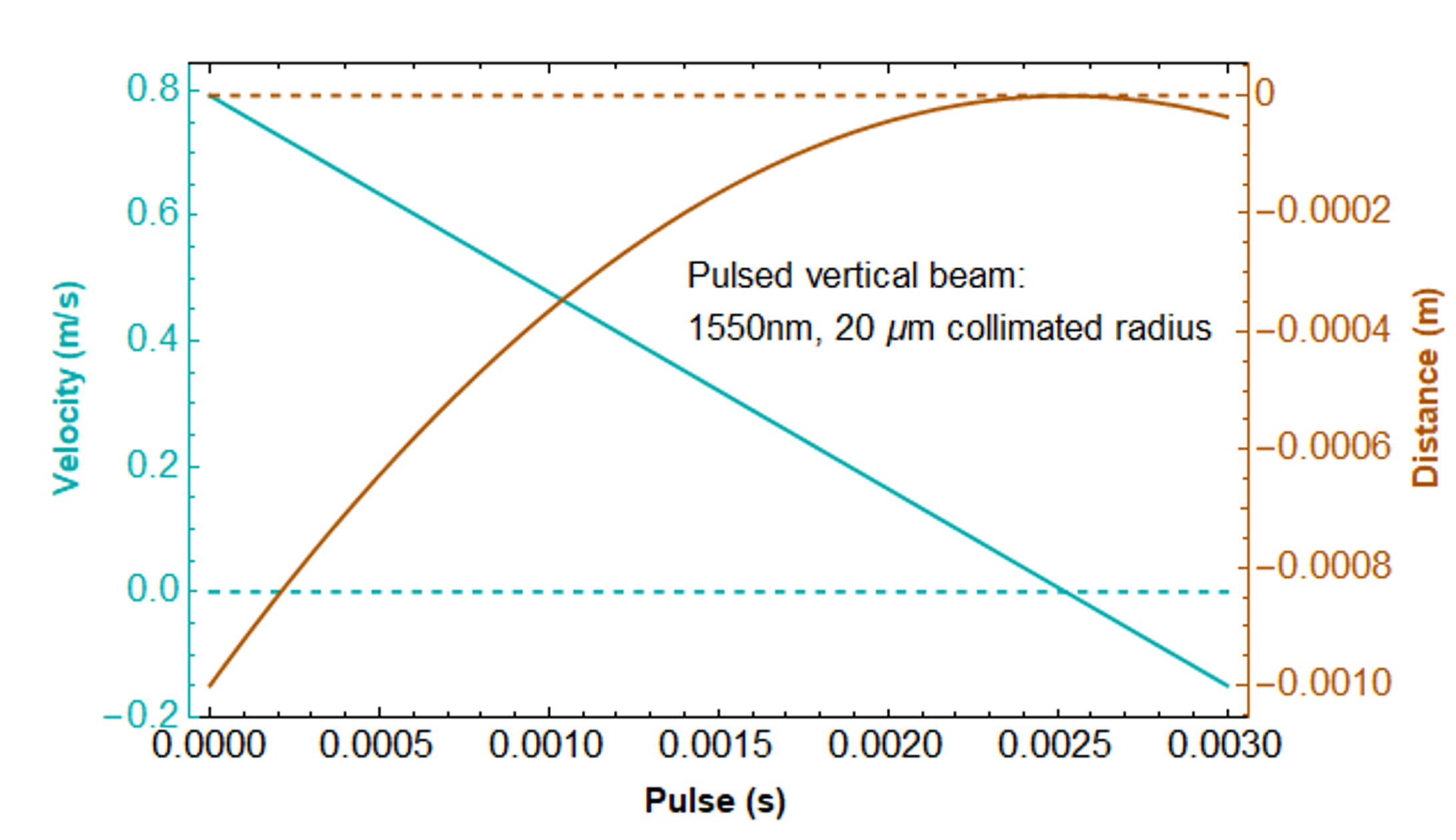}
	\caption{Velocity and distance vs slowing beam pulse duration. Cyan: Velocity of particles after falling 30 mm and then being slowed by a vertical beam pulse. Brown: Particle distance from the trap starting at 1 mm. Particle is stopped at the trap after 2.5 ms. As in Fig. \ref{fig:slow}, the slowing beam power for each sphere diameter is adjusted to achieve the same slowing trajectory: 170 nm at 5.36 W, 300 nm at 1 W, and 3 $\mu$m at .55 W}
	\label{fig:pulse}
\end{figure}

A $1064$ nm dual beam trap with $8 \mu$m waists, foci offset of $75 \mu$m and total power of 2.2 W yields $v = 16$ cm/s for a silica particle with $\rho$ = 2.2 kg/m$^3$ and $\epsilon$ = 2. Fig. \ref{fig:capvs} shows the velocities over a range of waists and several laser powers. A nanoparticle will be captured if its speed can be reduced below the velocities in Fig. \ref{fig:capvs} before it escapes radially. Trapping is achievable at moderate vacuum where collisions with gas molecules provide the necessary speed reduction. 
In high vacuum the particle will fall unimpeded and be pulled through the trap potential escaping capture. A focused vertical beam with sufficient power can then be implemented to slow the particle during free-fall and also reduce its velocity below the trap's capture velocity. Taking the dual beam trap configuration mentioned previously with an 8 $\mu$m waist, we calculate the velocities of 170 nm, 300 nm, and 3$\mu$m diameter particles as they fall from rest 33 mm above the trap in the presence of a focused vertical slowing beam that is focused to 20 $\mu$m at the trap center. The velocity of the particles are found from the total work:  
\begin{equation}
    \frac{1}{2}m v^2 = -\Delta U+m g \Delta r-\int F_s \,dr.
\end{equation}
 Here $m$ is the mass of a silica sphere, $\Delta U$ is the  change in trap potential along the radial or $r$ direction, $mg \Delta r$ is the change in gravitational potential, and 
 \begin{equation}\label{miescat}F_s = \frac{I}{c} C_{s}
 \end{equation}
 is the scattering force on a sphere of radius $a$ due to the vertical slowing beam of intensity $I$ and wavelength $\lambda$. $C_s = \frac{\lambda^2}{2 \pi} \Sigma_{n =1}^{\infty}(2n+1)(|a_n|^2+|b_n|^2)$ is the scattering cross section with Mie coefficients $a_n$ and $b_n$ \cite{kerker2013scattering}. For particle size smaller than the wavelength such that $\frac{2\pi}{\lambda} a << 1 $, equation \ref{miescat} reduces to the Rayleigh scattering result $F_s \approx \frac{128 \pi^5 a^6}{3 c \lambda^4}\left(\frac{(\epsilon-1)}{(\epsilon+2)}\right)^2I$ \cite{Harada}.  
 Fig \ref{fig:slow}. shows the velocity of the particles as they fall to the center of the trap ($\Delta r$ goes from -0.033 m to 0). The slowing beam parameters were chosen to slow the particles such that they don't exceed the capture velocity of the trap, but don't turn around (have negative velocity) before reaching the influence of the trap ($\sim$ 10 $\mu$m). The same trajectory is accomplished for particle diameters of 170 nm, 300 nm, and 3$\mu$m by adjusting the slowing beam power to 4.32 W, .79 W, and .448 W respectively.

 \begin{figure}
    \includegraphics[width=\linewidth]{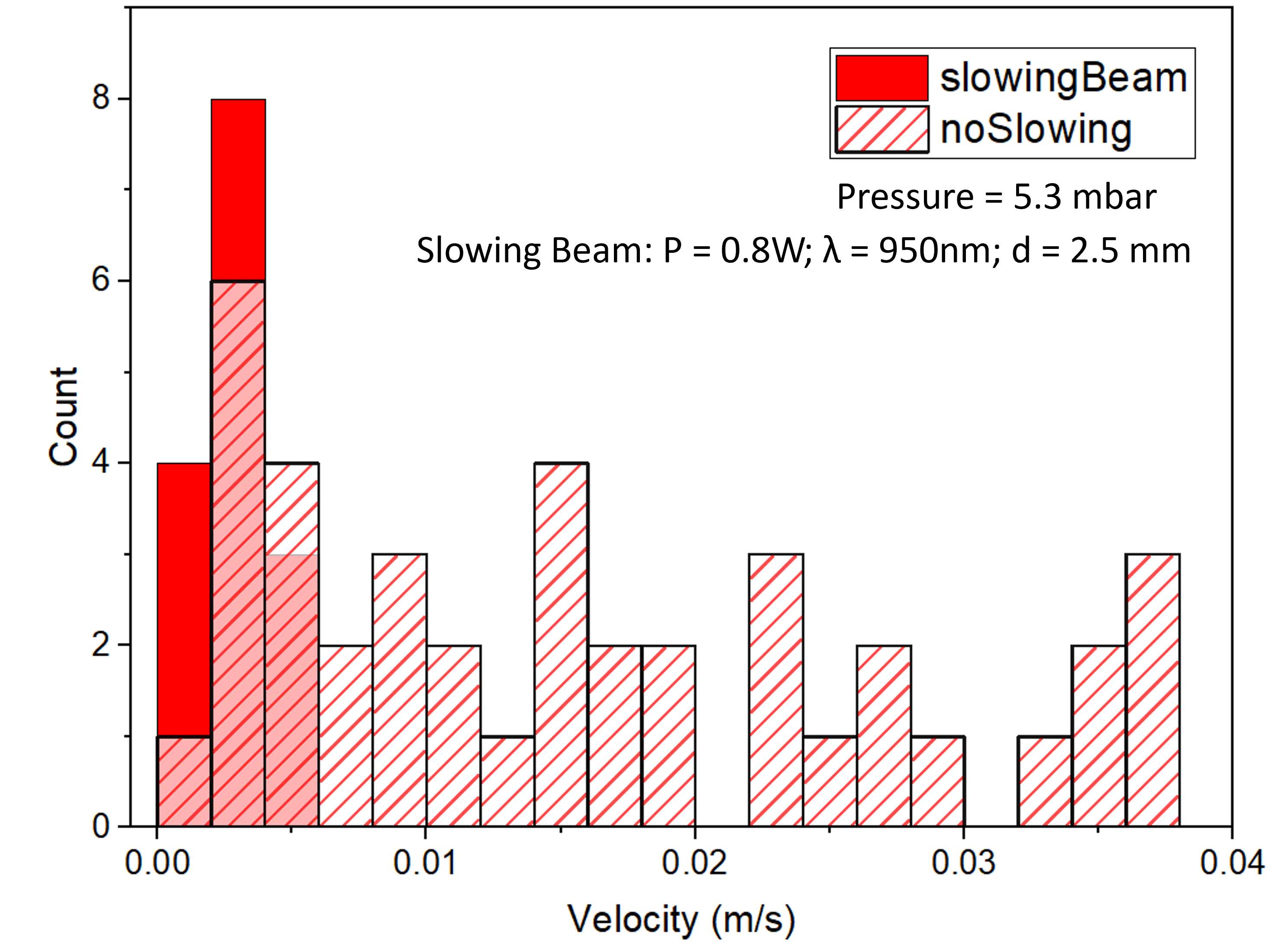}
	\caption{Velocity distribution of 170 nm particles launched with and without a vertical slowing beam at 5.3 mbar. The 950 nm wavelength slowing beam's power was 0.8 W and collimated to a 2.5 mm diameter. During these tests a clump of $170$ nm particles was observed to reverse direction before reaching the trap region (not shown in the histogram).}
	\label{fig:slowdata}
\end{figure}

Alternatively, a pulsed slowing beam triggered as the particles reach the trap could also be used to momentarily apply the scattering force to slow the particles enough to remain trapped. In our experiment at high vacuum, the particles reaches within 1 mm of the trap in about 0.08 seconds, achieving an instantaneous velocity of 0.79 m/s. A pulsed beam, for example collimated to a 20 $\mu$m radius with about 1 W of power, triggered 0.08 seconds after launch will stop a 300 nm particle that is 1 mm away from the trap after a 2.5 ms pulse. The particle velocities and trajectories starting at 1 mm from the trap are shown in Fig. \ref{fig:pulse} as they are slowed by a pulsed beam. The slowing beam power is again adjusted so that the particle diameters of 170 nm at 5.36 W, 300 nm at 1 W, and 3 $\mu$m at .55 W all achieve the same slowing trajectory.



   Initial tests using a collimated vertical beam were performed to demonstrate the slowing of 170 nm diameter particles. The particles were launched in the presence of the vertical beam at a moderate vacuum of 5.3 mbar towards a single beam trap about 30 mm below the launcher slide. The distribution of velocities with and without the slowing beam are shown in Fig. \ref{fig:slowdata}. along with the slowing beam parameters. Further details of the single beam setup used to trap 170 nm diameter particles as well as an expanded histogram of particle velocities at various pressures can be found in the appendix.

In contrast to the frame counting method described in section \ref{sec:velocity} to measure velocities, the speeds shown in Fig. \ref{fig:slowdata} were measured by tracking the particle's pixel location at three different frames, noting the time difference, and averaging the speeds. This method was only used to produce Fig. \ref{fig:slowdata}. since the vertical beam illuminated the particles for the entire duration of the fall from the launcher. 


\section{Discussion}
We described and characterized an in-vacuum and dry method for loading nanoparticles into optical traps. The high power pulsed ultrasonic piezoelectric launcher achieves  accelerations of $\sim 10^8$ m/s$^2$, large enough to release nanoparticles as small as 170 nm from the stiction and Van der Waals forces between the silica sphere and glass substrate. 
The wide distribution of 3 $\mu$m particle velocities at each pressure indicate the prevalence of clumps of particles that fall much more quickly than that expected for single particles. Snapshots and a qualitative description of the clump distributions launched in low and high vacuum can be found in Fig. \ref{fig:clumppics} of the appendix. 
Despite the presence of clumps, the particles accounted for in the velocity measurements of Fig. \ref{fig:VvsP} are on average consistent with 3 $\mu$m sized particles falling with zero initial velocity. Faster particles are likely filtered out from these measurements due to the angled launcher and only closely focusing on the particles falling through the trap region. Given the average speeds at high vacuum, it is feasible that a vertical beam can be used to slow the falling particles below the capture velocity. Trapping at high vacuum would be a boon for a variety experiments requiring short duty cycles such as nanoparticle interferometry.

\appendix*
\newcommand{\hbAppendixPrefix}{A}
\renewcommand{\thefigure}{\hbAppendixPrefix\arabic{figure}}
\setcounter{figure}{0}
\section{}

\begin{figure}
    \includegraphics[width=\linewidth]{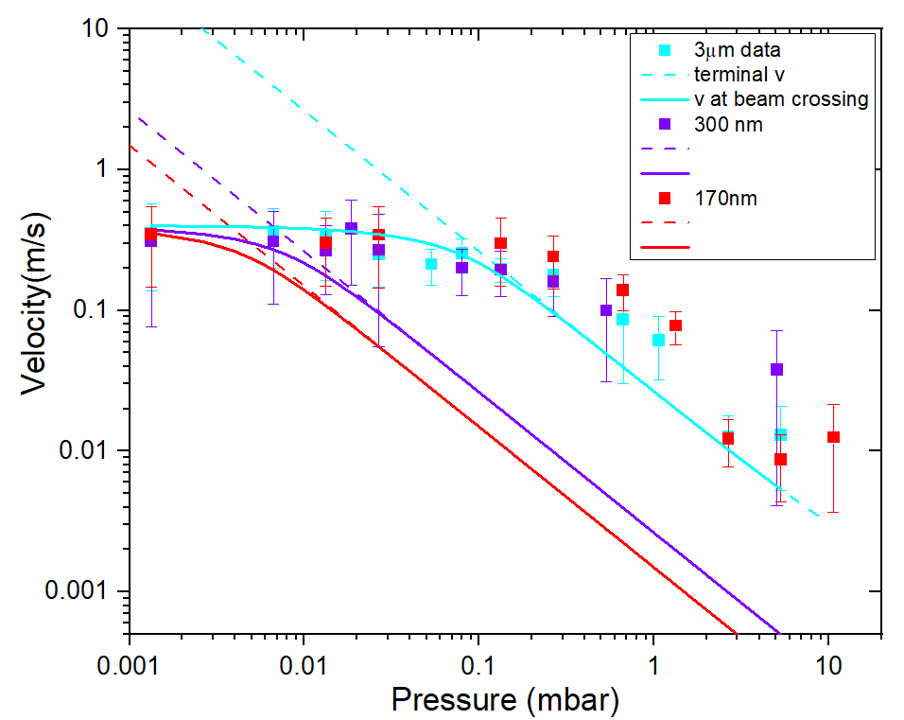}
	\caption{Velocity measurements for 3 $\mu$m, 300 nm, and 170 nm diameter particles. The velocity data for the 300 nm and 170 nm sized particles are well above their projected calculated velocities (solid lines) before reaching their expected velocity at high vacuum. This indicates that the vast majority of particles observed are large clumps of the smaller particle diameters.}
	\label{fig:VvsP_3sizes}
\end{figure}

\begin{figure}
    \includegraphics[width=.8\linewidth]{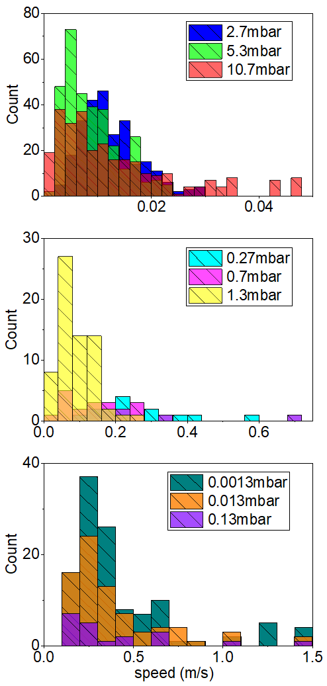}
	\caption{Histograms of 170 nm particle average velocities illuminated after traveling 30 mm to the focused 1596 nm beam trapping region for pressures ranging from $0.0013$ to $10.7$ mbar.The launching mechanism, detection camera and analysis method are the same used in the collection of data appearing in Fig. \ref{fig:VvsP}.}
	\label{fig:170nm_hists}
\end{figure}

\begin{figure}
    \includegraphics[width=\linewidth]{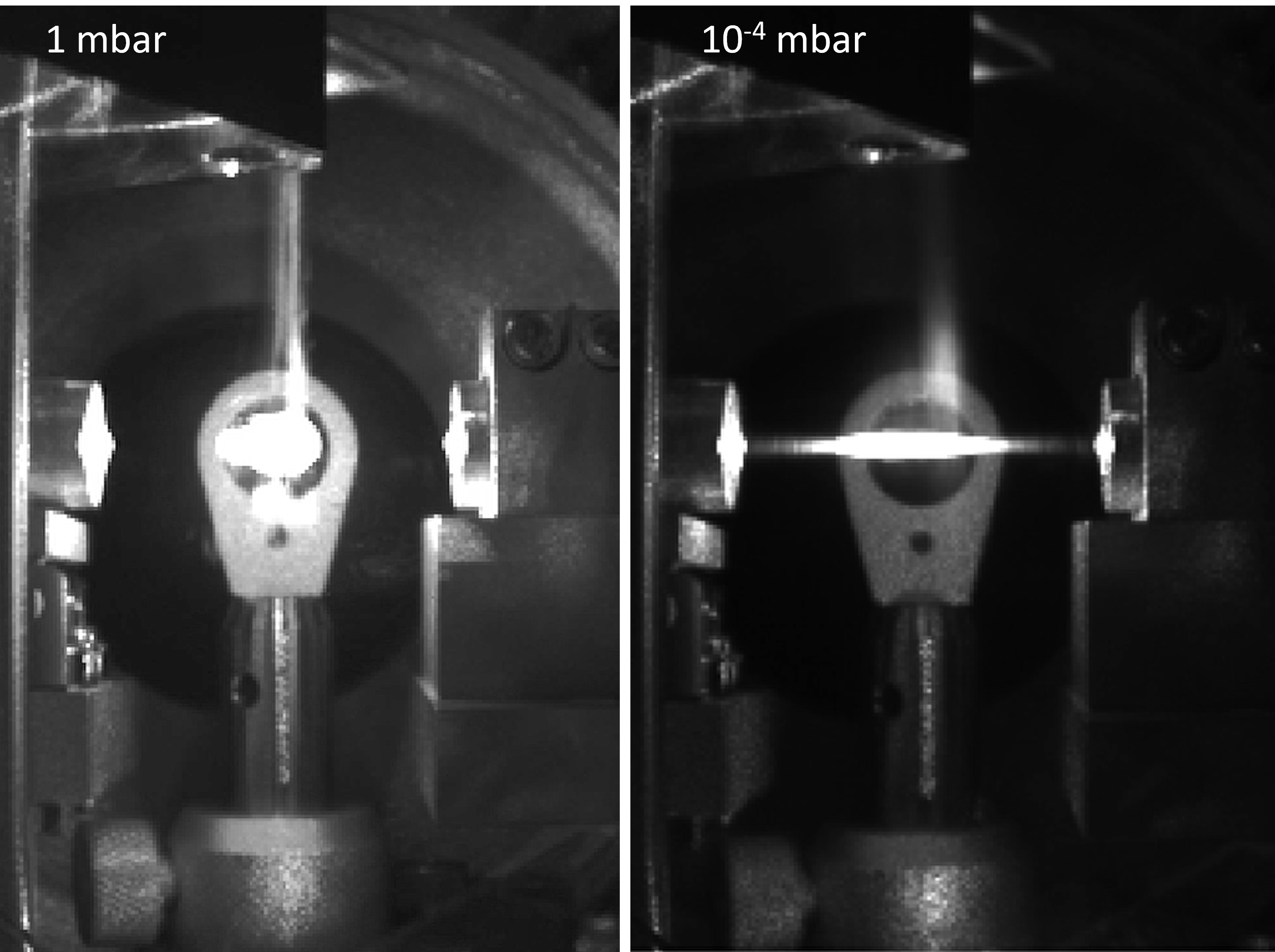}
	\caption{During launch (Left pic) at moderate vacuum $\sim$ 1 mbar there is stream of particles with an initial dense cloud of particle clumps at the head that quickly fall through the trap. This is followed by smaller more dispersed clumps of particles, and finally random small clumps and single particles wander through the trap. (Right pic) At high vacuum $\sim$ 10$^{-4}$ mbar the aforementioned vertical size spread is much less pronounced and all the particles almost fall at once through the trap.}
	\label{fig:clumppics}
\end{figure}

Additional velocity measurements were performed with 300 nm and 170 nm diameter particles. The 300 nm particles were launched in the same system described in section \ref{sec:velocity}, while a separate system with a 1596 nm single beam trap focused down to 1.6 $\mu$m \cite{montoya2022scanning} was used to measure the velocities of the 170 nm particles. The velocities however were measured in a similar manner as described in the main text. Fig.\ref{fig:VvsP_3sizes} shows the additional velocity data with all particle sizes and Fig. \ref{fig:170nm_hists} is the expanded histogram for 170 nm particles.  The velocity data for the 300 nm and 170 nm sized particles indicate speeds well above their projected calculated velocities that would be limited by terminal velocity. This indicates that a large fraction of emitted particles must consist of large clumps of the smaller particle diameters. This is consistent with findings of observed particles that have been launched from the substrated and collected for imaging in a scanning electron microscope. Both single spheres and clusters of size 170 nm, 300 nm, and 3 $\mu$ spheres are observed to be released from the glass substrates in all cases studied. Furthermore the larger clusters tend to scatter more light, making their observation of their trajectory in the CCD camera easier than for single particles.

A flat launcher (glass slide parallel to the trap) was used to observe the contents released from a glass slide freshly coated with 3 $\mu$m particles. Fig. \ref{fig:clumppics} (a) and (b) are snapshots from a high speed camera of the initial particles launched in low and high vacuum respectively. During launch at moderate vacuum $\sim$ 1 mbar there is stream of particles with an initial dense cloud of particle clumps at the head that quickly fall through the trap. This is followed by smaller more dispersed clumps of particles, and finally random small clumps and single particles wander through the trap.  At high vacuum $\sim$ 10$^{-4}$ mbar a large group of particles almost fall at once through the trap. After several launches the initial large cloud of clumps is almost entirely depleted.

\acknowledgements
We are grateful for discussions with T. Li, J. Millen, P. Barker, and H. Ulbricht. This work was partially supported by NSF grant nos. PHY-1505994, 1806686, the Heising-Simons Foundation, the J. Templeton Foundation, and the office of Naval Research grant no.417315//N00014-18-1-2370.

\bibliographystyle{unsrt}
\bibliography{bib}
\end{document}